\documentclass[%
 prl,
 mph,%
 amsmath,amssymb,
 reprint,%
]{revtex4-2}

\usepackage{graphicx}% Include figure files
\usepackage{dcolumn}% Align table columns on decimal point
\usepackage{bm}% bold math

\usepackage[mathlines]{lineno}% Enable numbering of text and display math
\modulolinenumbers[5]% Line numbers with a gap of 5 lines
\begin{document}

\preprint{PRL/123-QED}

\title[Curvature effect]{The Curvature Effect in Gaussian Random Fields}% Force line breaks with \\
%\thanks{Footnote to title of article.}

\author{Alexandre L. M. Levada}
 \email{alexandre.levada@ufscar.br}
\affiliation{ 
Computing Departament, Federal University of S\~ao Carlos, S\~ao Carlos, SP, Brazil%\\This line break forced with \textbackslash\textbackslash
}%

\date{\today}% It is always \today, today,

\begin{abstract}
Random field models are mathematical structures used in the study of stochastic complex systems. In this paper, we compute the shape operator of Gaussian random field manifolds using the first and second fundamental forms (Fisher information matrices). Using Markov Chain Monte Carlo techniques, we simulate the dynamics of these random fields and compute the Gaussian curvature of the parametric space, analyzing how this quantity changes along phase transitions. During the simulation, we have observed an unexpected phenomenon that we called the \emph{curvature effect}, which indicates that a highly asymmetric geometric deformation happens in the underlying parametric space when there are significant increase/decrease in the system's entropy. This asymmetric pattern relates to the emergence of hysteresis, leading to an intrinsic arrow of time along the dynamics.
\end{abstract}

\keywords{Gaussian random fields, information geometry, Fisher information, shape operator, curvature, entropy}
%Use showkeys class option if keyword %display desired
\maketitle

%\section{Introduction}
%\label{sec:Intro}

The dynamics of stochastic complex systems has recently garnered a lot of interest in the physics literature \cite{Baryam,Paolo}, directly contributing to the solution of several problems in social \cite{SocialNet}, biological \cite{MRFBiol} and economic \cite{MRFEcon} sciences. A relevant aspect concerns the prediction of phase transitions in a quantitative way by means of an objective mathematical criterion \cite{RandomFieldsTrans,PhaseIsing4D}. In this letter, we compute intrinsic geometric properties from the underlying parametric space of random fields composed by Gaussian variables \cite{GaussianRandomFields} and study how these quantities change along phase transitions.

Geometrodynamics is a research field whose main goal is to characterize and describe physical phenomena completely in terms of geometry \cite{Wheeler}, in an attempt to unify the fundamental forces and reformulate general relativity in terms of metric tensors of Riemannian manifolds \cite{Geometro}. These issues have been investigated by several physicists and remain an active field in the 21st century, as a mathematical tool for the unification of gravitation \cite{Geom2014}, quantum mechanics \cite{Unification} and in the study if particle systems \cite{Nature}. Information geometry a research field that combines information theory and Riemannian geometry to study intrinsic geometric properties of parametric spaces associated with random variables \cite{Amari,Frieden2004,Nielsen}. Inspired by these ideas, our study can be thought as an experimental attempt measure how the variation of the curvature in the underlying manifold (parametric space) of random field models is related to variations in system's entropy.

%In this context, our study can be thought as an experimental attempt to apply information geometry to completely characterize the dynamics of Gaussian random fields in terms of intrinsic properties of their parametric spaces  

At the beginning of the dynamics, the inverse temperature parameter is zero, and the random field model degenerates to a regular Gaussian distribution (independent random variables). In this situation, the parametric space exhibit constant negative curvature (hyperbolic geometry) \cite{Dodson}. The idea is to analyze how the emergence of a spatial dependence structure along time leads to irreversible geometric transformations in the parametric space. The obtained results show the existence of the \emph{curvature effect}, which can be described as: the variations of the Gaussian curvature when the system moves towards higher entropy states is different from the variations of the Gaussian curvature when the system moves towards lower entropy states, which induces the emergence of an intrinsic arrow of time as a natural one-way direction of evolution. The main objective of this scientific investigation is to propose an information-geometric framework to understand and characterize the dynamics of random fields defined on 2D lattices.

We assume some simplifying hypothesis: first, the random field is Markovian  (conditional independence principle). Second, the model is isotropic in the sense that the inverse temperature parameter, which control the spatial dependence structure, is spatially invariant and constant for all orientations in space. Last, but not least, we deal with a pairwise interaction model, which means that we allow only binary relationships. In summary, we consider a pairwise isotropic Gaussian-Markov random field to model the interaction between spatially dependent Gaussian random variables organized in a 2D lattice. The degree of interaction is quantified by a single coupling parameter, also known as the inverse temperature. With this model, it is possible to derive closed-form expressions for several expected values, which allows the exact computation of information-theoretic measures, such as Fisher information and entropy. Due to the Hammersley-Clifford theorem \cite{Hammersley}, we avoid the joint Gibbs distribution by defining a pairwise isotropic GMRF model through the set of local conditional density functions:

\begin{equation}
\footnotesize
	p\left( x_{i} | \eta_{i}, \vec{\theta} \right) = \frac{1}{\sqrt{2\pi\sigma^2}}exp\left\{-\frac{1}{2\sigma^{2}} \left[ x_{i} - \mu - \beta \sum_{j \in \eta_{i}} \left( x_{j} - \mu \right) \right]^{2} \right\}
	\label{eq:GMRF}
\end{equation} where $\eta_i$ denotes the second-order neighborhood system of $x_i$, $\vec{\theta} = (\mu, \sigma^{2}, \beta)$ denotes the vector of model parameters, where $\beta$ is the inverse temperature, which encodes the spatial dependence between the variables in the field.

Let $M$ be the underlying parametric space of our random field model. Then, $M$ is equipped with two fundamental forms. The first fundamental form (first-order Fisher information matrix), denoted by $I$, is the metric tensor, which allows the computation of inner products in the tangent planes. The second fundamental form (second-order Fisher information matrix), denoted by $II$, is composed by second order derivatives and quantifies how the manifold moves away from the tangent space at a given point. The shape operator of $M$, denoted by $P = -(II)(I)^{-1}$, encodes information about the curvature the manifold, being a powerful mathematical tool for geometric analysis \cite{Manfredo}. It has been shown that: 1) the Gaussian curvature, $K$, is the determinant of the shape operator $P$; 2) the mean curvature, $H$, is the trace of the shape operator $P$; and 3) the principal curvatures are the eigenvalues of the shape operator $P$. According to our mathematical derivations, the first-order Fisher information matrix (first fundamental form) of a pairwise isotropic GMRF is given by (for the complete mathematical derivation, please check \cite{Levada}):

\begin{equation}
	I(\vec{\theta}) = \left( \begin{array}{ccc}
	A & 0 & 0 \\ 
	0 & B & D \\ 
	0 & D & C
	\end{array} \right)
\end{equation} where 

\begin{widetext}
\begin{align}
	A & = \frac{\left(1 - \beta\Delta \right)^2}{\sigma^2} \left[ 1 - \frac{1}{\sigma^2}\left(  2\beta\sum_{j\in\eta_i}\sigma_{ij} - \beta^2\sum_{j\in\eta_i}\sum_{k\in\eta_i}\sigma_{jk} \right) \right] \\
	B & = \frac{1}{2\sigma^4} - \frac{1}{\sigma^6}\left[ 2\beta\sum_{j\in\eta_i}\sigma_{ij} - \beta^2 \sum_{j\in\eta_i}\sum_{k\in\eta_i}\sigma_{jk} \right] \\ \nonumber & + \frac{1}{\sigma^8}\left[ 3\beta^2 \sum_{j\in\eta_i}\sum_{k\in\eta_i}\sigma_{ij}\sigma_{ik} - \beta^3 \sum_{j\in\eta_i}\sum_{k\in\eta_i}\sum_{l\in\eta_i}\left( \sigma_{ij}\sigma_{kl} + \sigma_{ik}\sigma_{jl} + \sigma_{il}\sigma_{jk} \right) \right. \\ \nonumber & \hspace{1cm} \left. + \beta^4 \sum_{j\in\eta_i}\sum_{k\in\eta_i}\sum_{l\in\eta_i}\sum_{m\in\eta_i}\left( \sigma_{jk} \sigma_{lm} + \sigma_{jl}\sigma_{km} + \sigma_{jm}\sigma_{kl} \right)  \right] \\
	C & = \frac{1}{\sigma^2}\sum_{j\in\eta_i} \sum_{k\in\eta_i} \sigma_{jk} + \frac{1}{\sigma^4} \left[ 2 \sum_{j\in\eta_i} \sum_{k\in\eta_i} \sigma_{ij} \sigma_{ik} - 2\beta \sum_{j\in\eta_i} \sum_{k\in\eta_i} \sum_{l\in\eta_i} \left( \sigma_{ij}\sigma_{kl} + \sigma_{ik}\sigma_{jl} + \sigma_{il}\sigma_{jk} \right) \right. \\ \nonumber & \left. \qquad\qquad\qquad\qquad + \beta^2 \sum_{j\in\eta_i} \sum_{k\in\eta_i} \sum_{l\in\eta_i} \sum_{m\in\eta_i} \left( \sigma_{jk}\sigma_{lm} + \sigma_{jl}\sigma_{km} + \sigma_{jm}\sigma_{kl} \right) \right] \\
	D & = \frac{1}{\sigma^4}\left[ \sum_{j\in\eta_i}\sigma_{ij} - \beta\sum_{j\in\eta_i}\sum_{k\in\eta_i}\sigma_{jk} \right] \\ \nonumber & - \frac{1}{2\sigma^6}\left[ 6\beta\sum_{j\in\eta_i}\sum_{k\in\eta_i}\sigma_{ij}\sigma_{ik} - 3 \beta^2 \sum_{j\in\eta_i}\sum_{k\in\eta_i}\sum_{l\in\eta_i}\left( \sigma_{ij}\sigma_{kl} + \sigma_{ik}\sigma_{jl} + \sigma_{il}\sigma_{jk} \right) \right. \\ \nonumber & \hspace{3cm} \left. + \beta^3 \sum_{j\in\eta_i}\sum_{k\in\eta_i}\sum_{l\in\eta_i}\sum_{m\in\eta_i} \left( \sigma_{jk}\sigma_{lm} + \sigma_{jl}\sigma_{km} + \sigma_{jm}\sigma_{kl} \right) \right]
\end{align}
\end{widetext}

Similarly, the second-order Fisher information matrix (second fundamental form) is given by:

\begin{equation}
	II(\vec{\theta}) = \left( \begin{array}{ccc}
	K & 0 & 0 \\ 
	0 & L & N \\ 
	0 & N & M
	\end{array} \right)
\end{equation} where

\begin{align}
	K & = \frac{1}{\sigma^2}\left( 1 - \beta\Delta \right)^2 \\
	L & = \frac{1}{2\sigma^4} - \frac{1}{\sigma^6}\left[ 2\beta\sum_{j\in\eta_i} \sigma_{ij} - \beta^2 \sum_{j\in\eta_i}\sum_{k\in\eta_i}\sigma_{jk} \right] \\
	M & = \frac{1}{\sigma^2}\sum_{j\in\eta_i}\sum_{k\in\eta_i}\sigma_{jk} \\
	N & = \frac{1}{\sigma^4}\left[ \sum_{j\in\eta_i}\sigma_{ij} - \beta \sum_{j\in\eta_i}\sum_{k\in\eta_i}\sigma_{jk} \right]
\end{align}

Note that the elements of $I$ and $II$ are functions of the variance and the covariancies between the variables $x_i$ and $x_j$ in the random field belonging to the same second-order neighborhood system $\eta_i$, where $\Delta = 8$. 

Entropy is one of the most ubiquitous concepts in science, with applications in a large number of research fields. The entropy of a pairwise isotropic GMRF is:

\begin{align}
	\label{eq:entropia1}
	 H_{\beta}(\vec{\theta}) & = - E\left[ log~p\left(x_{i}| \eta_{i}, \vec{\theta} \right) \right] \\ \nonumber & = \frac{1}{2}\left[ log\left( 2\pi\sigma^2 \right) + 1\right] \\ \nonumber & \qquad\qquad - \frac{1}{\sigma^2} \left[ \beta\sum_{j \in \eta_i}\sigma_{ij} - \frac{\beta^2}{2}\sum_{j \in \eta_i}\sum_{k \in \eta_i}\sigma_{jk} \right] \\ \nonumber & = H_{G}(\vec{\theta}) - \beta \sigma^2 N - \beta^2\frac{M}{2}
\end{align} where $H_G(\vec{\theta})$ denotes the entropy of a Gaussian random variable and $M$ and $N$ are the components of the second fundamental form. Note that, as expected, for $\beta=0$, we have $H_{\beta}(\vec{\theta}) = H_{G}(\vec{\theta})$.

\begin{figure}[!h]
	\begin{center}
	\includegraphics[scale=0.23]{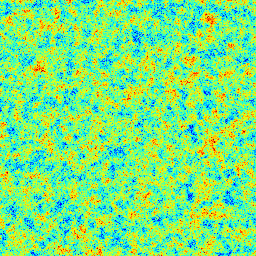}
	\includegraphics[scale=0.23]{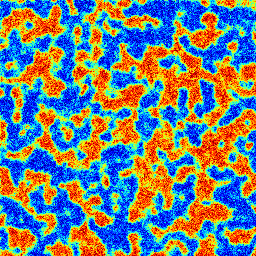}
	\includegraphics[scale=0.23]{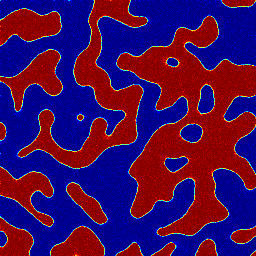}
	\includegraphics[scale=0.23]{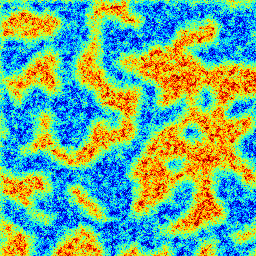}
	\end{center}
	\caption{Outcomes of the random field model along the MCMC dynamics.}
	\label{fig:MCMC}
\end{figure}

In order to perform computational simulations, we employed the Metropolis-Hastings algorithm. A full cycle of the dynamics is composed by 1000 iterations of the Markov-Chain Monte Carlo simulation: the inverse temperature parameter $\beta$ is initialized with zero, and at the end of each iteration we perform an infinitesinal displacement with $\beta$ being incremented by $\Delta\beta = 0.0006$, up to the $500$th iteration. In the second half of the cycle, at the end of each iteration, $\beta$ is decremented by $\Delta\beta = 0.0006$, until it reaches zero once again. The parameters $\mu$ and $\sigma^2$ are estimated by the sample mean and sample variance. Figure \ref{fig:MCMC} shows some outcomes of the random field during the evolution of the system.

\begin{figure}[!h]
	\begin{center}
	\includegraphics[scale=0.55]{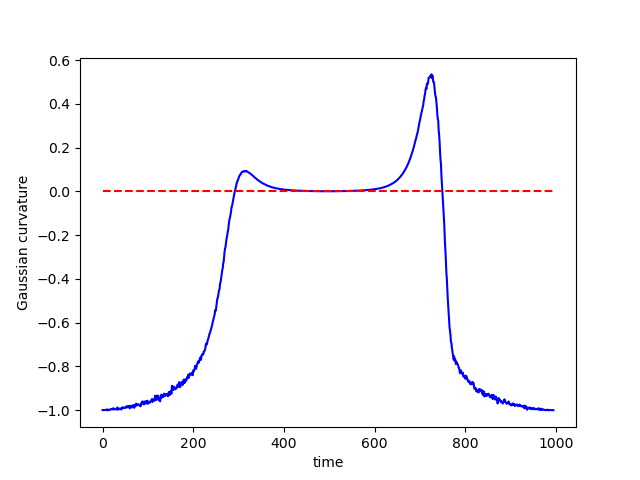}
	\end{center}
	\caption{Evolution of the Gaussian curvature along a full cycle of the MCMC simulation with the Gaussian random field model.}
	\label{fig:GC}
\end{figure}

During the first half of the dynamics, the entropy is increasing and the sign of the Gaussian curvature changes from negative to positive, whereas in the second half, the entropy decreases by the same amount and the sign of the Gaussian curvature changes from positive to negative. However, it is worth to mention that the obtained results show a non-intuitive phenomenon, named here as the \emph{curvature effect}. Basically, the variation of the Gaussian curvature is not symmetric during the dynamics. In the first phase transition, where the system's entropy drastically increases up to the maximum value, the curvature is smaller than that observed in the second phase transition, where the system's entropy drastically decreases down to the minimum value, even knowing that the variations in entropy are exactly the same. In other words, the amount of curvature necessary to bend and stretch/shrink the parametric space when moving towards lower entropy states is significantly higher than that necessary to bend and stretch/shrink the parametric space when moving towards higher entropy states. Figure \ref{fig:GC} illustrates the \emph{curvature effect} as an asymmetric pattern of evolution of the Gaussian curvature along the random field dynamics.

\begin{figure}[!h]
	\begin{center}
	\includegraphics[scale=0.55]{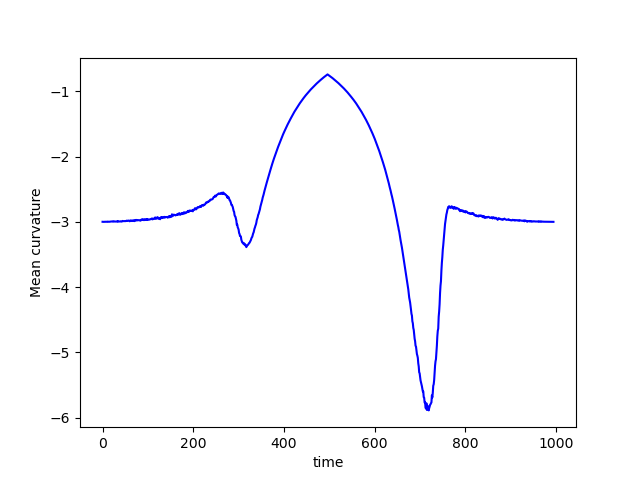}
	\end{center}
	\caption{Evolution of the mean curvature along a full cycle of the MCMC dynamics with the random field model.}
	%\caption{Evolution of the mean (M) and principal curvatures (K1, K2 and K3) along a full cycle of the MCMC simulation with the Gaussian random field model.}
	\label{fig:MC}
\end{figure}

It has been shown in differential geometry that the mean curvature is the summation of the principal curvatures and the Gaussian curvature is the product of the principal curvatures \cite{Manfredo}. Figure \ref{fig:MC} illustrates the variation of the mean curvature along the dynamics. Note that its variation is also highly asymmetric, indicating that, geometrically, the process of increasing entropy is significantly different from the process of decreasing entropy.

\begin{figure}[!h]
	\begin{center}
	\includegraphics[scale=0.5]{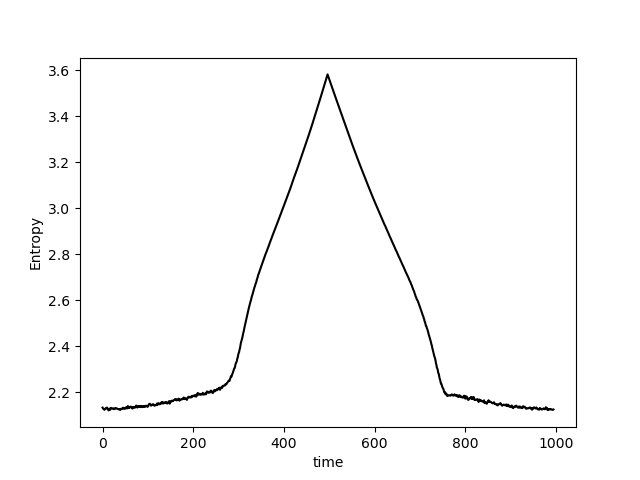}
	\end{center}
	\caption{Evolution of the system's entropy along a full cycle of the MCMC dynamics with the Gaussian random field model.}
	\label{fig:Entropy}
\end{figure}

Figure \ref{fig:Entropy} shows that the system's entropy increases in the first half of the dynamics and decreases by the same amount in the second half. It is possible to observe from the curves that the points of change in the sign of the Gaussian curvature coincides with the point in which the system's entropy has an abrupt change in its behavior.

\begin{figure}[!h]
	\begin{center}
	\includegraphics[scale=0.55]{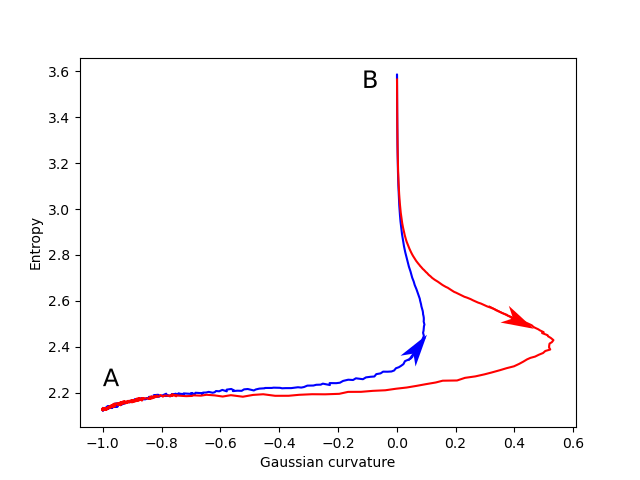}
	\end{center}
	\caption{The amount of stretching/shrinking the parametric space of Gaussian random fields suffers when it bends depends whether the entropy is increasing or decreasing, resembling a mathematical model of hysteresis.}
	\label{fig:curvature_path}
\end{figure}

Figure \ref{fig:curvature_path} shows how the system's entropy change as a function of the Gaussian curvature along a full cycle of the MCMC dynamics. The resulting behavior resembles a mathematical model of hysteresis, which underlies a large number of phase transitions in physical models \cite{HystIsing,Hysteresis}. The resulting pattern indicates that, due to the \emph{curvature effect}, the path from A (low entropy state) to B (high entropy state) is significantly smaller than the path from B to A, suggesting that the deformations in the parametric space needed to reduce the system's entropy by $\Delta H$ are greater than those needed to increase the system's entropy by the same $\Delta H$.

Note that this behavior induces a natural orientation to the process of, from a low entropy state, bringing the random field to a high entropy state and back, which can be considered an intrinsic arrow of time, represented by the variation of the curvature in the parametric space caused by changes in the inverse temperature parameter. 

Future works may include a deeper study about the relationship between curvature and the geodesic distances between two random fields operating in different regimes, as a way to provide an intrinsic similarity measure. We also intend to investigate techniques for the estimation of the inverse temperature parameter in order to simulate a situation in which we do not have direct access to the real inverse temperature value. Furthermore, we intend to study the curvature effect in other random field models, such as the classic Ising model and the q-state Potts models.

%, in which each variable assumes one of $q$ different discrete states.

%Another idea consists in computing information-theoretic divergences, such as the KL-divergence, between pairs of random field models in machine learning applications.

The intrinsic geometric structure of independent random variables has been extensively studied in information geometry. However, little is known about the geometry of complex systems modeled by random fields, where the inverse temperature parameter induces a spatial dependence structure. In this letter, we investigated how the variation of this parameter is related to the variation of the curvature of the manifolds of Gaussian random fields. The obtained results show that the variations in the mean and Gaussian curvatures are highly asymmetric, suggesting that the parametric space suffers a series of irreversible geometric deformations along the dynamics. Our geometric analysis has shown an unreported phenomenon: the \emph{curvature effect}, which suggests that the deformations in the parametric space are more prominent during a decrease of the system's entropy than during an increase of the system's entropy, suggesting the emergence of an intrinsic arrow of time in the system's dynamics. For the interested readers, the Python source code with the computational implementation of the numerical simulations are available at: \emph{https://github.com/alexandrelevada/Curvature\_GMRF}.

\begin{acknowledgments}
This study was financed in part by the Coordenação de Aperfeiçoamento de Pessoal de Nível Superior - Brasil (CAPES) - Finance Code 001
\end{acknowledgments}

\nocite{*}
\bibliography{aapmsamp}% Produces the bibliography via BibTeX.

\end{document}